\documentclass[twocolumn,preprintnumbers,showpacs,pre,floatfix,amsmath,amssymb,notitlepage,superscriptaddress]{revtex4-1}
\usepackage{epsfig}
\usepackage{amssymb}
\usepackage{bbm}
\usepackage{indentfirst}
\usepackage{graphicx}% Include figure files
\usepackage{amsmath,amssymb}
\usepackage{bpchem}

\usepackage{color}
\usepackage{tabularx}
\usepackage{multirow}
\usepackage{makecell}
\usepackage{array}
\usepackage{appendix}
\newcommand{\PreserveBackslash}[1]{\let\temp=\\#1\let\\=\temp}
\newcolumntype{C}[1]{>{\PreserveBackslash\centering}p{#1}}
\newcolumntype{R}[1]{>{\PreserveBackslash\raggedleft}p{#1}}
\newcolumntype{L}[1]{>{\PreserveBackslash\raggedright}p{#1}}

\input epsf

\begin{document}

\title{Spectrum of extensive multiclusters in the Kuramoto model with higher-order interactions}
\author{Can Xu}\email[]{xucan@hqu.edu.cn}
\affiliation{Institute of Systems Science and College of Information Science and Engineering, Huaqiao University, Xiamen 361021, China}

\author{Per Sebastian Skardal}
\email[]{persebastian.skardal@trincoll.edu}
\affiliation{Department of Mathematics, Trinity College, Hartford, Connecticut 06106, USA}

%\date{\today}

\newcommand{\WARN}[1]{\textcolor{green}{#1}}
\newcommand{\NOTES}[1]{\textcolor{red}{#1}}
\begin{abstract}
Globally coupled ensembles of phase oscillators serve as useful tools for modeling synchronization and collective behavior in a variety of applications. As interest in the effects of simplicial interactions (i.e., non-additive, higher-order interactions between three or more units) continues to grow we study an extension of the Kuramoto model where oscillators are coupled via three-way interactions that exhibits novel dynamical properties including clustering, multistability, and abrupt desynchronization transitions. Here we provide a rigorous description of the stability of various multicluster states by studying their spectral properties in the thermodynamic limit. Not unlike the classical Kuramoto model, a natural frequency distribution with infinite support  yields a population of drifting oscillators, which in turn guarantees that a portion of the spectrum is located on the imaginary axes, resulting in neutrally stable or unstable solutions. On the other hand, a natural frequency distribution with finite support allows for a fully phase-locked state, whose spectrum is real and may be linearly stable or unstable.
\end{abstract}

\pacs{05.45.Xt, 89.75.Fb, 89.75.-k}

\maketitle

\section{Introduction}\label{sec:01}

The emergence of synchronization in large populations of interacting units is one of the most well-known cooperative phenomena across a number of disciplines~\cite{strogatz2003sync,Pikovsky2003}. Such collective behaviors play central roles for functional relations in a wealth of different system including pacemaker cells in the heart~\cite{Glass1988}, Josephson junctions~\cite{benz1991coh}, power grids~\cite{Rohden2012PRL}, cell circuits~\cite{prindle2012sen}, and brain dynamics~\cite{buzski2004sci}. Exploring the intrinsic mechanism behind such self-organized behaviors has long been an important area of research that provides deep insights for understanding macroscopic dynamics in complex systems~\cite{Arenas2008PhysRep}.

One of the most useful paradigms for studying synchronization is the Kuramoto model, originally motivated by collective behavior in biological and chemical oscillators~\cite{kuramoto1975}, which consists of an ensemble of phase oscillators globally coupled via sinusoidal interaction and distributed natural frequencies. At a finite critical coupling strength the classical Kuramoto models exhibits an onset of synchrony that can be characterized by a phase transition from disorder to order and various types of coherent states towards synchronization can be observed~\cite{strogatz2000from,acebron2005the,pikovsky2015dyn}. In particular, the analytical tractability of the model has allowed researchers to better understand essential properties that give rise to collective, self-sustained oscillations. Moreover, by treating the transition to synchrony as a local bifurcation, the Kuramoto model provides a heuristic connection between nonlinear dynamics and statistical mechanics, which attracts increasing interest in uncovering the underlying mathematical and physical basis of the collective synchronization~\cite{ott2008low,ott2009long}.

Beyond the classical case, extensions of the Kuramoto model have been traditionally limited to pairwise interaction between oscillators, in which only the first-order harmonic of the phase difference in the coupling function is considered. However, recent work in physics and neuroscience applications highlights the potential importance of higher-order interactions described by non-pairwise, e.g., three-way or more, connections that may be organized via higher-order simplexes or a simplicial complex~\cite{Ashwin2016PhysD,Leon2019PRE,Petri2014Interface,Giusti2016JCN,Sizemore2018JCN,reimann2017cliques,Battiston2020PhysRep}. The effect of such interactions on the dynamics of complex systems therefore represents an important topic of research in the nonlinear dynamics community~\cite{komarov2015finite,bibk2016chaos,gong2019low,salnikov2019simplicial,skardal2011cluster,komarov2013multi,xu2016col,wang2017syn,filatrella2007gen,zou2020dyn,Millan2020PRL,Skardal2020,Skardal2020b,Lucas2020PRR}. A particularly interesting dynamical feature induced by higher-order interactions in systems of phase oscillators is the formation of multicluster states that display extensive multistability and a continuum of abrupt desynchronization transitions that allow oscilator systems to store memory and information~\cite{skardal2019abrupt}. The dynamical origin of such novel collective behavior was further addressed from a microscopic perspective~\cite{xu2020bif}. Despite these advances, a fundamental problem still lies in understanding how higher-order interactions give rise to the observed multicluster states and the structure of their corresponding spectral properties in the thermodynamic limit.

The aim of this paper is to provide detailed analysis of dynamical properties of multicluster states occurring in oscillator ensembles with higher-order coupling. We establish the continuity equation describing the macroscopic evolution of the coupled system, and identify corresponding steady states in terms of a parametrical self-consistent equation. More importantly, a complete description of the spectrum of the linearized evolution equation for fixed states is provided. In contrast to the case of the classical Kuramoto model, we prove rigorously that the drifting oscillators have no contribution to neither the order parameters nor the characteristic functions owing to the nonlinear higher-order coupling. Furthermore, we demonstrate that the continuous spectrum generated by the unperturbed linear operator does not contain any positive real part, and therefore can never induce instability. Based on the nontrivial roots of the the characteristic equations, we argue that the incoherent state, partially locked state, and fully locked state with arbitrary configuration are respectively neutral stable, linearly neutral stable, and linearly stable (unstable) to perturbations.

The remainder of this paper is organized as follows. In Sec.~\ref{sec:02} we present the governing equations, characterize their behaviors, and derive the parametrical self-consistent equations describing the steady states in the continuous limit. In Sec.~\ref{sec:03} we carry out a linear stability analysis of the multicluster state and characterize the properties of its spectrum in the thermodynamic limit. In Sec.~\ref{sec:04} we deduce the characteristic equations and discuss their roots to determine the eigenvalues of linearization. Finally, in Sec.~\ref{sec:05} we conclude with a discussion of our results.

\section{Model description and self-consistency equations}\label{sec:02}

We consider the dynamics of an ensemble of coupled phase oscillators with higher-order interactions (in particular, three-way interactions following \cite{skardal2019abrupt,Skardal2020b,xu2020bif}) that evolve according to
\begin{equation}\label{equ:01}
  \dot{\theta}_i=\omega_i+\frac{K}{N^2}\sum_{j=1}^{N}\sum_{k=1}^{N}\sin(\theta_j+\theta_k-2\theta_i),
\end{equation}
where $\theta_i\in S^1$ is the phase of oscillator $i$ ($i=1,\dots,N$), $\omega_i \in\mathbb{R}$ is its natural frequency chosen randomly from a distribution function $g(\omega)$, $N$ is the size of the system, and $K>0$ is a coupling strength among oscillators. In contrast to the standard Kuramoto model and its numerous generalizations, the interaction considered here is not pairwise but involves a triplet ($\theta_i, \theta_j, \theta_k$) that is equivalent to a fully connected hypernetwork topology. As we shall demonstrate below the hypernetwork organization of coupled oscillators results in a variety of nontrivial dynamical properties in both macroscopic and microscopic levels. Henceforth, unless explicitly noted, $g(\omega)$ is assumed to be an even function and is non-increasing for $\omega>0$ (symmetry and unimodal), while the support of $g(\omega)$ may be either finite or infinite.

Next, it is convenient to introduce two-complex valued order parameters to characterize the macroscopic dynamics of Eq.~(\ref{equ:01}) defined by
\begin{equation}\label{equ:02}
  Z_m=R_m e^{i \Theta_m}=\frac{1}{N}\sum_{j=1}^{N}e^{im\theta_j},\quad m=1, 2
\end{equation}
with $Z_m$ being the centroid of $N$ points $\{e^{im\theta_j}\}$ on the unit circle in the complex plane and $Z_1$ and $Z_2$ playing different roles. Usually, the Kuramoto order parameter $Z_1$ is sufficient to measure the synchrony provided that the coupling function does not include any higher-order Fourier modes. However, the interaction term presented here contains second order harmonics (i.e., $\sin[2(\theta_{jk}-\theta_i)]$, where $\theta_{jk}=(\theta_j+\theta_k)/2$), and it supports the formation of two clusters separated by phase difference $\pi$. Therefore, the Daido order parameter $Z_2$ is needed to quantify the overall degree of synchronization via entrainment in both clusters while $Z_1$ just measures the degree of asymmetry between the clusters. In particular, $R_1$ vanishes if the clusters are formed symmetrically and reaches a maximum when all entrained oscillator belong to the same cluster (leaving no oscillators in the opposite cluster).

\begin{figure*}[t]
\centering
\epsfig{file =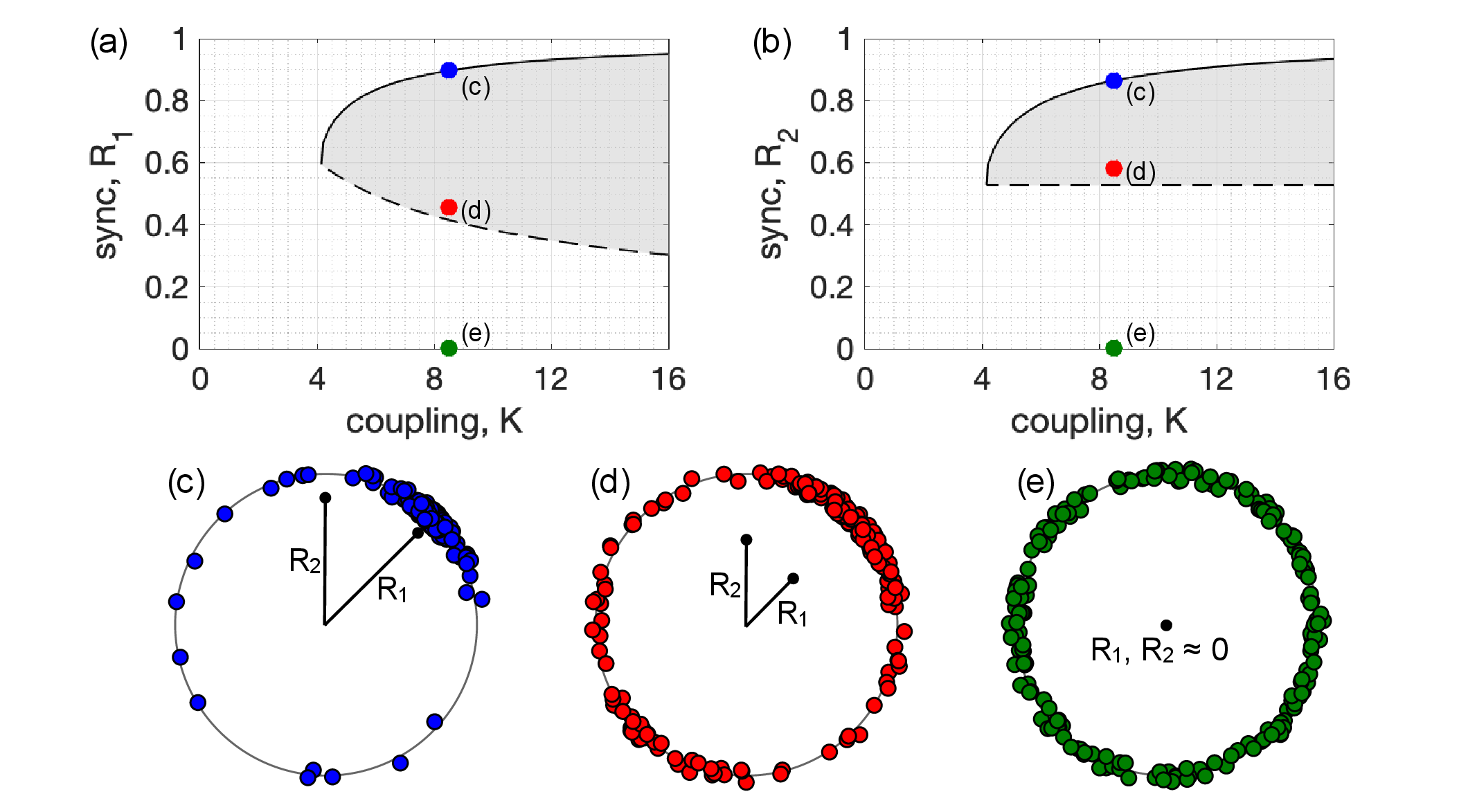, clip =,width=0.85\linewidth }
\caption{{\it Collective dynamics of the Kuramoto model with higher-order interactions.} For a case of $N=10^5$ oscillators with Lorentzian-distributed frequencies, $g(\omega)=[\pi(1+\omega^2)]^{-1}$ the possible order parameters (a) $R_1$ and (b) $R_2$ as a function of the coupling strength $K$ that correspond to stable multicluster states, denoted by the shaded areas. Below, examples of (c) a strongly asymmetric cluster state, (d) a more symmetric cluster state, and (e) the incoherent state, all of which are supported for $K=8.5$. (Of the $N=10^5$ oscillators, we depict $200$ chosen at random.)} \label{fig1}
\end{figure*}

To illustrate the different states that emerge from these dynamics we consider a system of $N=10^5$ oscillators with natural frequencies drawn from the Lorentzian distribution $g(\omega)=[\pi(1+\omega^2)]^{-1}$. In Fig.~\ref{fig1} we plot the possible values of the order parameters $R_1$ and $R_2$ as a function of the coupling strength $K$ that correspond to stable multicluster states, denoted by the shaded area, in panels (a) and (b). For both $R_1$ and $R_2$ the top branch (solid curve) represents the most asymmetric case where all entrained oscillators belong to the same cluster and the bottom branch (dashed curve) represents the most symmetric case supported by the corresponding coupling strength. This bottom branch represents a critical point under which multi cluster states lose stability in favor of the incoherent state. In panels (c), (d), and (e) we illustrate a strongly asymmetric cluster state, a more symmetric cluster state, and the incoherent state, respectively, all of which are supported for the choice of $K=8.5$, as a randomly chosen set of 200 oscillators. 

With a better understanding of the collective dynamics that emerge from Eq.~(\ref{equ:01}), we now move to derive the self-consistency equations. First, the definition of the order parameter allows us to rewrite Eq.~(\ref{equ:01}) in terms of the mean field:
\begin{equation}\label{equ:03}
  \dot{\theta}_i=\omega_i+KR_1^2\sin(2\Theta_1-2\theta_i).
\end{equation}
Thus, the evolution of each oscillator is solely coupled to $Z_1^2$, and $KR_1^2$ displays an effective force acting on it. We recall that the important property of Eq.~(\ref{equ:01}) is its rotation and reflection symmetry. Specifically, the dynamical equation remains unchanged under the transformation $\theta_i\rightarrow\theta_i+\alpha$ ($\alpha$ is a constant phase shift), and $(\theta_i, \omega_i)\rightarrow (-\theta_i, -\omega_i)$. This symmetry implies that for the long term evolution one can set $\Theta_m$ in Eq.~(\ref{equ:02}) to be zero by going into a rotating frame and shifted initial conditions.

Before proceeding with analysis, it is necessary to define an auxiliary parameter $q=KR_1^2$ and two particular inner products, namely,
\begin{equation}\label{equ:04}
  \langle \varphi, \psi\rangle=\int_{0}^{2\pi}\varphi \psi d\theta
\end{equation}
and
\begin{equation}\label{equ:05}
  (\varphi, \psi)=\int_{-\infty}^{\infty}\int_{0}^{2\pi}\varphi \psi g(\omega)d\theta d\omega,
\end{equation}
where $\varphi(\theta, \omega)$ and $\psi(\theta, \omega)$ are the integrable functions.

Next we consider the thermodynamical limit $N\rightarrow\infty$, allowing state of the coupled system to be described by a probability density function $\rho(\theta, \omega, t)$ which gives the relative number of $\theta_j\approx \theta$ and $\omega_j\approx \omega$ for a fixed time $t$. In fact, $\rho$ is a $2\pi$-periodic function with respect to $\theta$ and satisfies normalization condition
\begin{equation}\label{equ:06}
  \langle\rho(\theta, \omega, t), 1\rangle=1.
\end{equation}
The conservation of oscillators' number implies the continuous equation of the form
\begin{equation}\label{equ:07}
  \frac{\partial \rho}{\partial t}+\mathbf{D}(\rho v_\omega)=0,
\end{equation}
where $\mathbf{D}$ is differential operator on $S^1$ defined by the rule
\begin{equation}\label{equ:10}
  (\varphi, \mathbf{D}\psi)=(-\varphi', \psi),
\end{equation}
and ``$'$'' denotes derivation with respect to $\theta$ variable. The velocity field is given by
\begin{equation}\label{equ:08}
  v_\omega=\omega-q\sin 2\theta,
\end{equation}
and the order parameters in the continuous limit are evaluated as
\begin{equation}\label{equ:09}
  Z_m=(e^{im\theta}, \rho).
\end{equation}

The key step for the self-consistent analysis is to determine the stationary solutions of the continuity equation, such that $R_m$ remains constant. To this end, the oscillators can be divided into two groups, i.e., the phase locked and drifting oscillators, according to whether Eq.~(\ref{equ:03}) has a fixed point or not. For the phase locked oscillators, we have $|\omega|<q$ and $v_\omega(\theta_\omega^*)=0$, yielding
\begin{equation}\label{equ:11}
  \sin 2\theta_\omega^*=\frac{\omega}{q}\quad \mathrm{and}\quad \cos 2\theta_\omega^*=\sqrt{1-\omega^2/q^2}.
\end{equation}
Importantly, the higher-order coupling indicates that there exists two stable fixed points $\theta_\omega^*=\arcsin(\omega/q)/2$ and $\theta_\omega^*+\pi$, with the eventual behavior of each phase locked oscillator depending on its initial phase. For convenience, we denote the probability for the occurrence of $\theta_\omega^*$ and $\theta_\omega^*+\pi$ by $s_\omega$ and $1-s_\omega$, respectively. The stationary distribution formed by the phase locked oscillators is then given by
\begin{equation}\label{equ:12}
  \rho_\omega(\theta)=s_\omega\delta(\theta-\theta_\omega^*)+(1-s_\omega)\delta(\theta-\theta_\omega^*-\pi),
\end{equation}
where we have used $\delta(\cdot)$ to denote the Dirac function (unit point mass) and $s_\omega$ is the indicator function representing the ratio of oscillators belonging to the two clusters. For instance, if $s_\omega=1$, $\rho_\omega(\theta)$ corresponds to the unit point mass in the cluster near $\theta=0$, and $s_\omega<1$ measures the deviation from this state with an emerging cluster near $\theta=\pi$. Similar to $g(\omega)$, we restrict ourselves to unimodal and symmetric case of $s_\omega$. Furthermore, trigonometric identities imply that
\begin{equation}\label{equ:13}
  \cos\theta_\omega^*=\sqrt{\frac{1+\sqrt{1-\omega^2/q^2}}{2}},
\end{equation}
and
\begin{equation}\label{equ:14}
  \sin\theta_\omega^*=\frac{\omega/q}{2\sqrt{(1+\sqrt{1-\omega^2/q^2})/2}}.
\end{equation}

On the other hand, drifting oscillators satisfy $|\omega|>q$ and therefore do not reach a fixed point. For a stationary state to emerge we require that $\rho v_\omega$ is constant, which gives
\begin{equation}\label{equ:15}
  \rho_\omega(\theta)=\frac{C_\omega}{v_\omega},
\end{equation}
with normalization constant
\begin{equation}\label{equ:16}
  C_\omega^{-1}=\int_{0}^{2\pi}v_\omega^{-1}(\theta)d\theta,
\end{equation}
which is an odd function of $\omega$. Different from the phase locked scenario, the drifting oscillators do not belong to any cluster, each rotating non-uniformly on the unit circle with periodic $T_\omega=|C_\omega^{-1}|$, thereby forming a smooth probability measure in the thermodynamic limit.

Turning to the order parameters, they are evaluated as
\begin{equation}\label{equ:17}
  Z_m=(e^{im\theta}, \rho_\omega)_{|\omega|<q}+(e^{im\theta},\rho_\omega)_{|\omega|>q},
\end{equation}
where we have split the integral into two parts, and each part is defined in a symmetrical interval. The first part corresponds to the phase locked oscillators and the second corresponds to the drifting ones. Note that, the second integral vanishes since $C_\omega$ is an odd function implying that the drifting oscillators have no contribution to $Z_m$. Likewise, the imaginary part of the first integral is zero due to the fact that $\sin m\theta_\omega^*$ is an odd function of $\omega$. These vanishing integrals in turn imply that the order parameters $Z_m$ are real -- a property that coincides with our previous assumption of $\Theta_m=0$.

Substituting Eqs.~(\ref{equ:12}-\ref{equ:14}) into Eq.~(\ref{equ:17}) leads to
\begin{equation}\label{equ:18}
  Z_m=R_m=\langle s_\omega,\cos m\theta_\omega^*\rangle+\langle 1-s_\omega,\cos m(\theta_\omega^* +\pi)\rangle.
\end{equation}
Using that $q=KR_1^2$, the self-consistent equations may be rewritten
\begin{equation}\label{equ:19}
  R_2=G(q)=\int_{-q}^{q}\sqrt{1-\omega^2/q^2}g(\omega)d\omega,
\end{equation}
and
\begin{equation}\label{equ:20}
  \frac{1}{\sqrt{K}}=F(q)=\int_{-q}^{q}(2s_\omega-1)\sqrt{\frac{1+\sqrt{1-\omega^2/q^2}}{2q}}g(\omega)d\omega.
\end{equation}

The self-consistent equations (\ref{equ:19}-\ref{equ:20}) form a close description of the stationary solutions of the Kuramoto-type system of globally coupled phase oscillators with higher-order interactions. In addition to the incoherent state ($R_m=0$), which exists for any value of $K$ and remains neutrally stable to perturbation in $N\rightarrow \infty$~\cite{strogatz1991stability}, Eqs.~(\ref{equ:19}-\ref{equ:20}) define the curves describing the steady-state solutions representing multicluster states of Eq.~(\ref{equ:01}) in the half plane of the parameters $K>0$ and $q>0$. BBy varying $q$ from 0 to $\infty$, we obtain a parametric representation of synchronization transition of the form $(K, R_1, R_2)=(F^{-2}(q), qF^2(q), G(q))$. We remark that Eq.~(\ref{equ:20}) defines a map from $q\in [0,\infty)$ to $K\in[0,\infty)$ through the function $F(q)$. However, since $F(q)$ is not necessary one to one, it may happen that for some parameter $K$ the inverse $F^{-1}(K)$ does not exist at all~\cite{omel2013bifurcation}. In other words, it is multivalued and has different solutions $q$ for the same parameter $K$. As a result, the underlying bifurcation takes plays in two different scenarios, either $F'(q_c)=0$ (smooth fold bifurcation) or $F'(q_c)$ does not exist (no smooth fold bifurcation). We denote $q_c$ to be the critical parameter such that $F(q_c)$ is a local extremum and the associated critical points for synchronization transition are
\begin{equation}\label{equ:21}
  (K_c, R_1^c, R_2^c)=(F^{-2}(q_c),q_cF^2(q_c),G(q_c)).
\end{equation}
Notice also that, besides $q_c$, there exists another critical point $q_l$ beyond which all oscillators are phase locked, such as $q_l=1$ when $g(\omega)$ has a finite support, and $q_l=\infty$ when $g(\omega)$ is defined in an infinite interval.

We demonstrate now that for a particular choice of the indicator function, we can observe non-universal synchronization transitions. For example, if $s_\omega$ is taken to be a constant $s$ ($1/2\leq s \leq 1$) and $g(\omega)$ is smooth, then the continuous function $F(q)$ is scaled as
\begin{equation}\label{equ:22}
\begin{split}
  &F(q)=(2s-1)f(q)\\
=&(2s-1)\int_{-q}^{q}\sqrt{\frac{1+\sqrt{1-\omega^2/q^2}}{2q}}g(\omega)d\omega.
\end{split}
\end{equation}
It is easy verified that $f(0)=f(\infty)=0$ and there is a unique $q_c\in(0,\infty)$ where $f(q_c)$ is a maximum. Given the stability of the incoherent state, the physical picture of phase transitions towards synchronization is illustrated as follows. (\romannumeral1) As the coupling strength $K$ between oscillators is increased adiabatically from zero, a spontaneous phase transition from the incoherence to synchrony can never be identified. (\romannumeral2) If the inverse process is adopted, beginning at a partially or fully locked state, the system stays at a partially or fully locked state, and there are an infinite number of branches of solutions with non-zero $R_m$ since $s\in[\frac{1}{2},1]$ that represent different multicluster states. Next, as $K$ decreases below a critical $K_c$, the extensive multiclusters disappear accompanied by a continuum of abrupt desynchronization transitions~\cite{komarov2015finite,skardal2019abrupt}, where the corresponding critical points for a given $s$ satisfy a scaling forming $(K_c, R_1^c, R_2^c)=([(2s-1)f(q_c)]^{-2}, q_c (2s-1)^2f^2(q_c), G(q_c))$. It should be pointed out that this nontrivial dynamical property essentially differs from conventionally first-order phase transition, where the order parameter undergoes a discontinuous jump for both forward and backward progresses and a bistability of incoherence and coherence coexist only in the hysteresis region.

\section{linear stability analysis of steady states}\label{sec:03}
The analysis above provides a framework for the complete description of the steady states in terms of the self-consistent equations with arbitrary $s_\omega$ and $g(\omega)$, which however can not address any stability properties of the corresponding solutions. In this section, we study the stability properties of the multicluster states formed by the nonlinear higher-order coupling. To this end, we preform a linear stability analysis of stationary distribution $\rho_\omega(\theta)$ with respect to continuity equation. In the following, we will show that the stability of $\rho_\omega(\theta)$ is reduced to the eigen-spectrum analysis of a linear operator which can be systematically worked out by a pair of characteristic equations.

To linearize Eq.~(\ref{equ:07}), we impose a small perturbation $\epsilon \eta_\omega(\theta, t)$ to $\rho_\omega(\theta)$ with $0<\epsilon\ll 1$. The perturbation $\eta_\omega(\theta, t)$ is a vector of tangent space $\mathbb{T}$ and satisfies the orthogonality condition,
\begin{equation}\label{equ:23}
  \langle\eta_\omega(\theta, t), 1\rangle=0,
\end{equation}
since $\rho(\theta, \omega, t)$ itself is normalized. Then the order parameters under the perturbation reduce to
\begin{equation}\label{equ:24}
\begin{split}
  Z_m[\eta_\omega]&=(e^{im\theta}, \rho_\omega+\epsilon \eta_\omega)\\
&=R_m+\epsilon (e^{im\theta},\eta_\omega).
\end{split}
\end{equation}
Inserting Eq.~(\ref{equ:24}) into the mean-field equation (\ref{equ:03}), the corresponding vector field becomes
\begin{equation}\label{equ:25}
\begin{split}
  v[\eta_\omega]=&\omega+KR_1^2 \sin(2\Theta_1-2\theta)\\
=&\omega+2K \mathrm{Im} Z_1[\eta_\omega] \mathrm{Re} Z_1[\eta_\omega] \cos2\theta \\
&-K((\mathrm{Re} Z_1[\eta_\omega])^2+(\mathrm{Im} Z_1[\eta_\omega])^2)\sin 2\theta,
\end{split}
\end{equation}
with
\begin{equation}\label{equ:26}
  \mathrm{Re} Z_1[\eta_\omega]=R_1+\epsilon(\cos\theta, \eta_\omega),
\end{equation}
\begin{equation}\label{equ:27}
  \mathrm{Im} Z_1[\eta_\omega]=\epsilon(\sin\theta, \eta_\omega).
\end{equation}
The first-order deviation of the velocity field corresponding to the tangent vector $\eta_\omega$ is equal to
\begin{equation}\label{equ:28}
  \epsilon v_1[\eta_\omega]=2\epsilon KR_1 (s_\eta \cos 2\theta- c_\eta \sin2 \theta),
\end{equation}
where we denote $c_\eta=(\cos\theta, \eta_\omega)$ and $s_\eta=(\sin\theta, \eta_\omega)$, respectively, as the perturbations of the real and imaginary part of the order parameter $Z_1$. Replacing $\rho_\omega$ with $\rho_\omega+\epsilon\eta_\omega$ and $v_\omega$ with $v_\omega+\epsilon v_1$ in Eq.~(\ref{equ:07}), we obtain to linear order in $\epsilon$ the equation
\begin{equation}\label{equ:29}
  \frac{d\eta_\omega}{dt}+\mathbf{D}(v_\omega\eta_\omega+v_1 \rho_\omega)=0.
\end{equation}
This leads to the definition of the linearized evolution equation at the fixed state $\rho_\omega$ as
\begin{equation}\label{equ:30}
  \frac{d\eta_\omega}{dt}=\mathcal{L}\eta_\omega,
\end{equation}
where $\mathcal{L}$ is a linear operator defined by
\begin{equation}\label{equ:31}
  \mathcal{L}\eta_\omega=-\mathbf{D}(v_\omega \eta_\omega)+2KR_1 \mathbf{D}((c_\eta \sin 2\theta -s_\eta \cos 2\theta)\rho_\omega)
\end{equation}
on the tangent space $\mathbb{T}$. We are now tasked with completely describing the spectrum of $\mathcal{L}$.

Following general spectrum theory of linear operators, we are interested in two parts of the spectrum $\sigma(\mathcal{L})$: the point (discrete) spectrum $\sigma_p(\mathcal{L})$, and continuous (essential) spectrum $\sigma_c(\mathcal{L})$. The point spectrum $\sigma_p(\mathcal{L})$ consisting of all eigenvalues $\lambda\in \mathbb{C}$ of $\mathcal{L}$, i.e., $\lambda$ values that make $(\lambda \mathbf{I}-\mathcal{L})$ not invertible and $\mathrm{ker}(\lambda \mathbf{I}-\mathcal{L})$ finite dimensional. Similarly, the continuous spectrum $\sigma_c(\mathcal{L})$ includes all $\lambda\in \mathbb{C}$ that make $(\lambda \mathbf{I}-\mathcal{L})$ not invertible, but rather $\mathrm{ker}(\lambda \mathbf{I}-\mathcal{L})=\{0\}$ and the image $\mathrm{Im}(\lambda \mathbf{I}-\mathcal{L})$ is unbounded. In fact, the elements of $\sigma_p(\mathcal{L})$ are similar to eigenvalues of a finite dimensional matrix, in that it corresponds to the finite dimensional degenerations of the operator $(\lambda \mathbf{I}-\mathcal{L})$. In contrast, $\sigma_c(\mathcal{L})$ represents the infinite dimensional nature of the operator $\mathcal{L}$.

To better understand the spectrum of $\mathcal{L}$, it is convenient to express $\mathcal{L}$ as
\begin{equation}\label{equ:31'}
  \mathcal{L}=\mathcal{M}+\mathcal{B},
\end{equation}
with the operators $\mathcal{M}$ and $\mathcal{B}$ defined by
\begin{equation}\label{equ:32}
  \mathcal{M}\eta_\omega=-\mathbf{D}(v_\omega \eta_\omega),
\end{equation}
and 
\begin{equation}\label{equ:33}
  \mathcal{B}\eta_\omega=2KR_1 \mathbf{D}((c_\eta \sin 2\theta -s_\eta \cos 2\theta)\rho_\omega).
\end{equation}
Remarkably, the rank of $\mathcal{B}$ is only two, since it has a codimensional-two kernel, determined by the conditions $s_\eta=c_\eta=0$. We can think of $\mathcal{B}$ as a perturbed operator with finite rank added to $\mathcal{M}$. In this sense, the operators $\mathcal{L}$ and $\mathcal{M}$ are quite similar.

Before analyzing the spectrum of $\mathcal{L}$, we split the tangent space as a direct sum of two subspaces $\mathbb{T}=\mathbb{T}_e\oplus \mathbb{T}_o$, where $\mathbb{T}_e$ and $\mathbb{T}_o$ denote the even and odd subspace, respectively. For a function, $\varphi(\theta, \omega)$ under the reflection action $\hat{\kappa}\varphi(\theta, \omega)=\varphi(-\theta, -\omega)$, if $\hat{\kappa}\varphi=\varphi$, then we call $\varphi$ even, or odd if $\hat{\kappa}\varphi=-\varphi$. For instance, $v_\omega(\theta)$ is odd, and $\rho_\omega(\theta)$ is even. Any vector $\eta\in \mathbb{T}$ can be decomposed as the sum of an even and an odd element vector uniquely. If $\eta_\omega(\theta)\in \mathbb{T}_e$, we have $s_\eta=0$, because the integral
\begin{equation}\label{equ:34}
\begin{split}
\int_0^{2\pi}\sin\theta\cdot\eta_\omega(\theta)d\theta&=\int_0^{2\pi}\sin\theta\cdot\eta_{-\omega}(-\theta)d\theta\\
&=\int_0^{2\pi}\sin(-\theta)\cdot\eta_{-\omega}(\theta)d\theta\\
&=-\int_0^{2\pi}\sin\theta \cdot\eta_{-\omega}(\theta)d\theta,
\end{split}
\end{equation}
is an odd function of $\omega$. Likewise, if $\eta_\omega(\theta)\in \mathbb{T}_o$, then $c_\eta=0$. This definition indicates that the even and odd tangent vectors correspond to pure cosine and sine perturbations of the order parameters, respectively. Notice that the operators $\mathcal{M}$, $\mathcal{B}$, as well as $\mathcal{L}$ all preserve the even and odd subspace. To see this, note that for an even $\eta_\omega$, $v_\omega\eta_\omega$ is odd, then $\mathbf{D}(v_\omega\eta_\omega)$ is even, and $\sin 2\theta\cdot \rho_\omega(\theta)$ is odd, so $\mathbf{D}(\sin 2\theta\cdot\rho_\omega(\theta))$ is even. Hence, $\mathcal{M}\eta_\omega\in\mathbb{T}_e$, and the proof is similar for odd vectors. Depending on whether $|\omega|<q$ or not, the subspace can be further split into the direct sum $\mathbb{T}_o=\mathbb{T}_o^l\oplus\mathbb{T}_o^d$ and $\mathbb{T}_e=\mathbb{T}_e^l\oplus\mathbb{T}_e^d$, where $\mathbb{T}_o^l (\mathbb{T}_e^l)$ and $\mathbb{T}_o^d (\mathbb{T}_e^d)$ are the subspaces of tangent vectors supported on the locked and drifting frequencies, respectively. This decomposition of the tangent space $\mathbb{T}$ allows us to analyze the spectrum of $\mathcal{L}$ by considering each subspace separately.

We begin with the spectrum of $\mathcal{M}$ and, as discussed above, it suffices to study $\mathcal{M}$ separately on the locked subspace $\mathbb{T}^l=\mathbb{T}_o^l\oplus\mathbb{T}_e^l$ and drifting subspace $\mathbb{T}^d=\mathbb{T}_o^d\oplus\mathbb{T}_e^d$, and then combine the results. For the distribution $\rho_\omega(\theta)$ in Eq.~(\ref{equ:12}) of the phase locked oscillators, the associated tangent vector $\eta_\omega(\theta)\in \mathbb{T}^l$ takes the following form~\cite{mirollo2007spe}
\begin{equation}\label{equ:35}
  \eta_\omega=a(\omega)\mathbf{D}\delta_{\theta_\omega^*}+b(\omega)\mathbf{D}\delta_{\theta_\omega^*+\pi},
\end{equation}
with coefficients $a(\omega)$ and $b(\omega)$. Assuming $\varphi$ to be a smooth function on $S^1$, then
\begin{equation}\label{equ:36}
\begin{split}
(\varphi, \mathcal{M}\eta_\omega)&=(\varphi, -\mathbf{D}(v_\omega\eta_\omega))\\
&=(v_\omega\varphi', a(\omega)\mathbf{D}\delta_{\theta_\omega^*})+(v_\omega\varphi', b(\omega)\mathbf{D}\delta_{\theta_\omega^*+\pi})\\
&=-a(\omega)(v_\omega\varphi')'_{\theta=\theta_\omega^*}-b(\omega)(v_\omega \varphi')'_{\theta=\theta_\omega^*+\pi}.
\end{split}
\end{equation}
Since $v_\omega(\theta_\omega^*)=v_\omega(\theta_\omega^*+\pi)=0$, so we have
\begin{equation}\label{equ:37}
  (\varphi,\mathcal{M}\eta_\omega)=-a(\omega)(v'_\omega\varphi')_{\theta=\theta_\omega^*}-b(\omega)(v'_\omega\varphi')_{\theta=\theta_\omega^*+\pi}.
\end{equation}
Hence,
\begin{equation}\label{equ:38}
\begin{split}
  \mathcal{M}\eta_\omega &=-2q\cos 2\theta_\omega^* \cdot (a(\omega)\mathbf{D}\delta_{\theta_\omega^*}+b(\omega)\mathbf{D}\delta_{\theta_\omega^*+\pi})\\
&=-2q\cos2\theta_\omega^*\cdot\eta_\omega.
\end{split}
\end{equation}
From this, we see that the application of the operator $\mathcal{M}$ on $\mathbb{T}^l$ turns out to be a multiplication by the function $-2q\cos 2\theta_\omega^*$. This implies that the spectrum of $\mathcal{M}$ on $\mathbb{T}^l$ is continuous, as $\omega$ ranges over the locked frequencies ($|\omega|<q$). Therefore, $\sigma(\mathcal{M})$ in $\mathbb{T}^l$ should be discussed in two different cases, if $q>q_l$ and $g(\omega)$ is supported on a finite interval $\rho_\omega(\theta)$ is fully locked, then $\sigma(\mathcal{M})=[-2q, -2\sqrt{q^2-1}]$. However, if $q<q_l$, $\rho_\omega(\theta)$ is partially phase locked, $\sigma(\mathcal{M})=[-2q, 0]$.

For the drifting oscillators ($|\omega|>q$) and $\eta_\omega\in \mathbb{T}^d$, the distribution $\rho_\omega(\theta)$ in Eq.~(\ref{equ:15}) has more regularity than $\rho_\omega(\theta)$ in Eq.~(\ref{equ:12}), and the calculations of $\sigma(\mathcal{M})$ are relatively easy. This is because
\begin{equation}\label{equ:39}
  (\varphi, \mathcal{M}\eta_\omega)=(\mathcal{M}^\dagger\varphi, \eta_\omega)=(v_\omega\varphi', \eta),
\end{equation}
where $\mathcal{M}^\dagger$ is adjoint operator of $\mathcal{M}$ on $\mathbb{T}^d$, which is defined as
\begin{equation}\label{equ:40}
  \mathcal{M}^\dagger \varphi=v_\omega \varphi'.
\end{equation}
Thus, $\lambda\in\sigma(\mathcal{M})$ on $\mathbb{T}^d$ is just the eigenvalue of $\mathcal{M}^\dagger$ which satisfies eigen-equation
\begin{equation}\label{equ:41}
  \lambda\varphi-v_\omega \varphi'=0,
\end{equation}
with $\varphi$ being a nontrivial solution in $S^1$. Solving it to obtain the general solution of $\varphi$
\begin{equation}\label{equ:42}
  \varphi=C \exp\{\lambda \int_0^\theta v_\omega^{-1}(x)dx\},
\end{equation}
where $C$ is a nonzero constant. Since $\varphi(\theta+2\pi)=\exp\{\lambda/C_\omega\}\varphi(\theta)$, the periodic condition for $\varphi$ requires $\exp\{\lambda/C_\omega\}=1$ leading to
\begin{equation}\label{equ:43}
  \lambda=2\pi i nC_\omega=in \,\mathrm{sgn}(\omega) \sqrt{\omega^2-q^2},
\end{equation}
for some $n\in \mathbb{Z}$. Therefore, $\mathcal{M}$ has only continuous spectrum on $\mathbb{T}$, $\sigma(\mathcal{M})=[-2q, -2\sqrt{q^2-1}]$ if $q>q_l$ and the system is fully phase locked, and $\sigma(\mathcal{M})=[-2q ,0]\cup i\mathbb{R}$ if $q<q_l$ and the system is partially phase locked.

We turn to the spectrum analysis of the full operator $\mathcal{L}$ on $\mathbb{T}$. As shown in ref.~\cite{mirollo2007spe}, $\sigma_c(\mathcal{L})=\sigma(\mathcal{M})$ due to the fact that the two operators $\mathcal{L}$ and $\mathcal{M}$ differ only by a bounded operator $\mathcal{B}$ of finite rank and the essential spectrum is invariant under finite rank perturbation. Thus, the next task is to determine $\sigma_p(\mathcal{L})$. As before, it suffices to study $\mathcal{L}$ separately on the even and odd subspaces. If $\eta_\omega\in \mathbb{T}_e$, $\mathcal{L}$ reduces to
\begin{equation}\label{equ:44}
  \mathcal{L}\eta_\omega=\mathcal{M}\eta_\omega+2KR_1 c_\eta \mathbf{D}(\sin 2\theta \cdot\rho_\omega).
\end{equation}
For a given $\lambda\in \mathbb{C}\setminus\sigma(\mathcal{M})$, if the operator $(\lambda \mathbf{I}-\mathcal{M})$ is invertible, the equation $\mathcal{L}\eta_\omega=\lambda\eta_\omega$ holds for $c_\eta\neq 0$, otherwise, $\lambda\in \sigma(\mathcal{M})$. We can assume $c_\eta=(2KR_1)^{-1}$ without loss of generality, then the eigenvector $\eta_\omega$ is determined by the following formula,
\begin{equation}\label{equ:45}
  \eta_\omega=(\lambda \mathbf{I}-\mathcal{M})^{-1}\mathbf{D}(\sin 2\theta\cdot\rho_\omega).
\end{equation}
Defining the function,
\begin{equation}\label{equ:46}
  h_c(\lambda)=(\cos\theta, (\lambda \mathbf{I}-\mathcal{M})^{-1}\mathbf{D}(\sin 2\theta\cdot\rho_\omega)),
\end{equation}
and according to the relation, $c_\eta= (2KR_1)^{-1}$, the eigenvalue $\lambda$ of $\mathcal{L}$ on $\mathbb{T}_e$ satisfies the self-consistent equation,
\begin{equation}\label{equ:47}
  \frac{1}{\sqrt{K}}=2\sqrt{q}h_c(\lambda), \; \lambda\in\mathbb{C}\setminus\sigma(\mathcal{M}).
\end{equation}
Similarly, if $\eta_\omega\in \mathbb{T}_o$, $\mathcal{L}$ is given by
\begin{equation}\label{equ:48}
  \mathcal{L}\eta_\omega=\mathcal{M}\eta_\omega-2KR_1 s_\eta\cdot \mathbf{D}(\cos 2\theta\cdot \rho_\omega)
\end{equation}
and $h_s(\lambda)$ is defined by
\begin{equation}\label{equ:49}
  h_s(\lambda)=-(\sin\theta, (\lambda\mathbf{I}-\mathcal{M})^{-1}\cdot \mathbf{D}(\cos2\theta\cdot \rho_\omega)).
\end{equation}
Imposing the relation $s_\eta=(2KR_1)^{-1}$, then the eigenvalue of $\mathcal{L}$ on $\mathbb{T}_o$ satisfies the self-consistent equation,
\begin{equation}\label{equ:50}
  \frac{1}{\sqrt{K}}=2\sqrt{q}h_s(\lambda), \; \lambda\in\mathbb{C}\setminus\sigma(\mathcal{M}).
\end{equation}
Here, $h_c(\lambda)$ and $h_s(\lambda)$ are the characteristic functions, and all $\lambda\in\sigma_p(\mathcal{L})$ correspond the roots of these characteristic equations (\ref{equ:47}) and (\ref{equ:50}). In the next section we turn to characterize these roots. 

\section{roots of the characteristic equations}\label{sec:04}
We now move to derive the explicit formulas for the characteristic equations, and then discuss their roots to determine the stability properties of multicluster states. As we did with the operator $\mathcal{M}$ above, we need to split the integrals in $h_c(\lambda)$ and $h_s(\lambda)$ [defined in Eqs.~(\ref{equ:46}) and (\ref{equ:49})] into the locked and drifting parts, namely,
\begin{equation}\label{equ:51}
  h_c(\lambda)= h_c^l(\lambda)+ h_c^d(\lambda)
\end{equation}
and
\begin{equation}\label{equ:52}
  h_s(\lambda)= h_s^l(\lambda)+ h_s^d(\lambda).
\end{equation}
Moving forward, the key step is to compute the expression $(\lambda\mathbf{I}-\mathcal{M})^{-1}\cdot \mathbf{D}(e^{2i\theta}\rho_\omega)$ explicitly.

We begin with the locked case $|\omega|<q$, and the operator $(\lambda\mathbf{I}-\mathcal{M})^{-1}$ degenerates to the multiplication by the function $\lambda+2q\cos 2\theta_\omega^*$. Hence, the contributions from the locked oscillators to the characteristic functions are
\begin{equation}\label{equ:53}
\begin{split}
  h_c^l(\lambda)&=(\cos \theta, (\lambda\mathbf{I}-\mathcal{M})^{-1}\mathbf{D}(\sin 2\theta\cdot\rho_\omega))\\
&=(\cos \theta, (\lambda+2q\cos2\theta_\omega^*)^{-1}\mathbf{D}(\sin 2\theta\cdot\rho_\omega))\\
&=(\sin \theta, (\lambda+2q\cos2\theta_\omega^*)^{-1}\sin 2\theta\cdot\rho_\omega)\\
&=\int_{-q}^q (2s_\omega -1)\frac{\sin\theta_\omega^* \sin 2\theta_\omega^*}{\lambda+2q\cos2\theta_\omega^*}g(\omega)d\omega,
\end{split}
\end{equation}
and
\begin{equation}\label{equ:53}
\begin{split}
  h_s^l(\lambda)&=(-\sin \theta, (\lambda\mathbf{I}-\mathcal{M})^{-1}\mathbf{D}(\cos 2\theta\cdot\rho_\omega))\\
&=(-\sin \theta, (\lambda+2q\cos2\theta_\omega^*)^{-1}\mathbf{D}(\cos 2\theta\cdot\rho_\omega))\\
&=(\cos \theta, (\lambda+2q\cos2\theta_\omega^*)^{-1}\cos 2\theta\cdot\rho_\omega)\\
&=\int_{-q}^q (2s_\omega -1)\frac{\cos\theta_\omega^* \cos 2\theta_\omega^*}{\lambda+2q\cos2\theta_\omega^*}g(\omega)d\omega.
\end{split}
\end{equation}
Remarkably, $h_c^l(\lambda)$ and $h_s^l(\lambda)$  are just the continuous limit ($N\rightarrow \infty$) of rational functions in refs.~\cite{xu2020bif,mirollo2005spe}.

Next, for the drifting case $|\omega|>q$, the distribution $\rho_\omega(\theta)$ is a smooth function and therefore $(\lambda\mathbf{I}-\mathcal{M})^{-1}\mathbf{D}(\sin 2\theta\cdot \rho_\omega)$ is a smooth function on $S^1$ of the form
\begin{equation}\label{equ:55}
  (\lambda\mathbf{I}-\mathcal{M})^{-1}\mathbf{D}(\sin 2\theta\cdot\rho_\omega)=\alpha_\omega(\theta),
\end{equation}
applying the operator $(\lambda\mathbf{I}-\mathcal{M})$ to both sides of Eq.~(\ref{equ:55}) and considering the definition of $\mathbf{D}$, $\alpha_\omega(\theta)$ must satisfy the following differential equation
\begin{equation}\label{equ:56}
  \lambda\alpha_\omega+(v_\omega \alpha_\omega)'=C_\omega(\frac{\sin 2\theta}{v_\omega})'.
\end{equation}
For convenience, we express $\alpha_\omega=\beta_\omega/v_\omega^2$ to clear out the denominator $v_\omega^2$ above, then the equation for $\beta_\omega$ is
\begin{equation}\label{equ:57}
  \lambda \beta_\omega-v'_\omega\beta_\omega+v_\omega\beta'_\omega=C_\omega(2\cos 2\theta\cdot v_\omega-\sin2\theta\cdot v'_\omega).
\end{equation}
Our strategy for solving this equation is to express $\beta_\omega$ in the form of Fourier decomposition as
\begin{equation}\label{equ:58}
  \beta_\omega(\theta)=\sum_{n=-\infty}^{\infty}c_n e^{in\theta},
\end{equation}
where the coefficients satisfy $c_{-n}=\bar{c}_n$ since $\beta_\omega(\theta)$ is real. Then the first order differential equation for $\beta_\omega(\theta)$ is equivalent to the algebraic equation of $c_n$ yielding
\begin{equation}\label{equ:59}
\begin{split}
  &[\lambda+q(e^{2i\theta}+e^{-2i\theta})]\sum_n c_n e^{in\theta}\\
&+[\omega-\frac{q}{2i}(e^{2i\theta}-e^{-2i\theta})]\sum_n in c_n e^{in\theta}\\
=&\omega C_\omega (e^{2i\theta}+e^{-2i\theta}).
\end{split}
\end{equation}
Balancing the coefficients of each Fourier mode, we obtain
\begin{equation}\label{equ:60}
  c_0=-\frac{4 C_\omega q\omega}{\lambda^2+4\omega^2-4q^2},
\end{equation}
while
\begin{equation}\label{equ:61}
  c_2=\frac{C_\omega\lambda\omega+2iC_\omega \omega^2}{\lambda^2+4\omega^2-4q^2},
\end{equation}
and
\begin{equation}\label{equ:62}
  c_{\pm1}=c_{\pm n}=0\quad (n>2).
\end{equation}
Then $\beta_\omega(\theta)$ takes the form
\begin{equation}\label{equ:63}
  \beta_\omega(\theta)=c_0+c_2 e^{2 i\theta}+c_{-2}e^{-2i\theta}.
\end{equation}
Furthermore, a straight forward calculation reveals the orthogonality relations $\langle e^{i\theta},v_\omega^{-2}\rangle=0$ and $\langle e^{i\theta}, e^{i2\theta}v_\omega^{-2}\rangle=0$, which in turn implies that $h_c^d(\lambda)=0$.

Following a similar process, we can express $(\lambda\mathbf{I}-\mathcal{M})^{-1}\mathbf{D}(\cos 2\theta\cdot\rho_\omega)$ as $\beta_\omega/v_\omega^2$ to obtain the evolution equation for $\beta_\omega$. The similar result holds where $\beta_\omega$ is a linear combination of  $(1, \sin2\theta, \cos 2\theta)$, therefore, we have $h_s^d(\lambda)=0$. We emphasize that the simple results originate from the nonlinear higher-order coupling in contrast to the case of the traditional Kuramoto-like models. We conclude that the drifting oscillators have no contribution to neither the order parameters nor the characteristic functions, which further verify our assumption in ref.~\cite{xu2020bif} rigorously. Finally, substituting $h_c(\lambda)=h_c^l(\lambda)$ and $h_s(\lambda)=h_s^l(\lambda)$ into Eqs.~(\ref{equ:47}) and (\ref{equ:50}), we arrive at a pair of characteristic equations describing the eigenvalues of $\lambda$, namely
\begin{equation}\label{equ:64}
\begin{split}
  \frac{1}{\sqrt{K}}=&2\sqrt{q}h_c(\lambda)\\
=&\sqrt{2}q^{-3/2}\int_{-q}^{q}(2s_\omega-1)\frac{\omega^2}{\lambda+2q\sqrt{q^2-\omega^2}}\\
&\cdot\frac{1}{\sqrt{1+\sqrt{1-\omega^2/q^2}}}g(\omega)d\omega,
\end{split}
\end{equation}
and
\begin{equation}\label{equ:65}
\begin{split}
  \frac{1}{\sqrt{K}}=&2\sqrt{q}h_s(\lambda)\\
=&\sqrt{2}q^{-1/2}\int_{-q}^{q}(2s_\omega-1)\frac{\sqrt{q^2-\omega^2}}{\lambda+2q\sqrt{q^2-\omega^2}}\\
&\cdot{\sqrt{1+\sqrt{1-\omega^2/q^2}}}g(\omega)d\omega,
\end{split}
\end{equation}

The final task is to determine the roots of the characteristic equations. Notice first that
\begin{equation}\label{equ:66}
  \frac{1}{\sqrt{K}}=2\sqrt{q}h_s(0).
\end{equation}
This means that $\lambda=0$ is always a root of the characteristic equation (\ref{equ:65}), and the property comes from the rotation symmetry of Eq.~(\ref{equ:01}), and the associated eigenvector corresponds to a trivial uniform perturbation of the system. Next, Eqs.~(\ref{equ:64}) and (\ref{equ:65}) show that $\lambda\in \mathbb{R}$. Otherwise, if $\lambda\in \mathbb{C}$ has nonzero imaginary part, the same is true for $h_c(\lambda)$ and $h_s(\lambda)$. To see this, we can multiply the numerator and denominator in the integral by $\bar{\lambda}+2\sqrt{q^2-\omega^2}$. Hence, the characteristic equations must have real roots. In addition, if $\lambda<-2q$, we have $h_c(\lambda)<0$ and $h_s(\lambda)<0$. So $\lambda<-2q$ can never be a root of the characteristic equations.

Let's consider the partially locked state $q<q_l$. In this case, the locked frequencies $\omega\in[-q, q]$ causing the continuous spectrum of $\mathcal{L}$ in the real axis to change from $-2q$ to $0$. Whereas, $h_c(\lambda)$ and $h_s(\lambda)$ are not defined for $-2q<\lambda<0$, since there are simple poles in the integrals of formulas. Therefore, any roots $\lambda<0$ for the characteristic equations are ruled out in the partially locked case. On the other hand, $h_c(\lambda)$ and $h_s(\lambda)$ are well defined and positive for $\lambda\geq 0$, and both functions are strictly decreasing on the interval $[0, \infty)$. So each characteristic equation has at most one root. We have already shown that Eq.~(\ref{equ:65}) has a trivial root $\lambda=0$, so it has no other roots.

Next, we claim that $2\sqrt{q}h_c(0)-1/\sqrt{K}<0$, which implies that the characteristic equation (\ref{equ:64}) has no roots at all. To see this, observe that
\begin{equation}\label{equ:67}
\begin{split}
  &2\sqrt{q}h_c(0)-\frac{1}{\sqrt{K}}\\
=&\sqrt{2q}\int_0^1 [2s(qx)-1]h(x)g(qx)dx,
\end{split}
\end{equation}
where $h(x)=\frac{x^2}{\sqrt{1-x^2}}\frac{1}{\sqrt{1+\sqrt{1-x^2}}}-\sqrt{1+\sqrt{1-x^2}}$ changes the sign from negative to positive at $x_0=\sqrt{3}/2$. $g(qx)$ and $s(qx)$ are non-increasing on $[0,1]$, then
\begin{equation}\label{equ:68}
\begin{split}
&2\sqrt{q}h_c(0)-\frac{1}{\sqrt{K}}\\
=&\sqrt{2q}\int_0^{x_0}[2s(qx)-1]h(x)g(qx)dx\\
&+\sqrt{2q}\int_{x_0}^1 [2s(qx)-1]h(x)g(qx)dx\\
\leq &\sqrt{2q}[2s(qx_0)-1]g(qx_0)\cdot \int_0^1 h(x)dx\\
=&-\frac{2}{3}\sqrt{2q}[2s(qx_0)-1]g(qx_0)<0.
\end{split}
\end{equation}
This inequality shows that $\mathcal{L}$ has only a trivial eigenvalue $\lambda=0$ in the partially synchronized state.

We turn to the fully locked state $q>q_l$ and prove that the characteristic equation (\ref{equ:64}) has a negative root if $F'(q)<0$ and the positive root if $F'(q)>0$. In this situation, $\omega\in[-1,1]$ in the integral of formulas, and $\sigma_c(\mathcal{L})=[-2q, -2\sqrt{q^2-1}]$  where the continuous spectrum of $\mathcal{L}$ is restricted to the real axis. Based on the same reason above, $\lambda$ is a root of the characteristic equations only if $\lambda>-2\sqrt{q^2-1}$. $h_c(\lambda)$ and $h_s(\lambda)$ are well defined and positive for $\lambda>-2\sqrt{q^2-1}$ and both functions tend to $+\infty$ if $\lambda\rightarrow -2\sqrt{q^2-1}$ from the right hand side. The decreasing property of $h_c(\lambda)$ and $h_s(\lambda)$ on the interval $(-2\sqrt{q^2-1}, \infty)$ ensures that the characteristic equations each can have at most one root. Also, $\lambda=0$ has been proven to be a trivial root of Eq.~(\ref{equ:65}), so we just need to consider Eq.~(\ref{equ:64}). Then Eq.~(\ref{equ:64}) has a negative root $-2\sqrt{q^2-1}<\lambda<0$ if and only if $2\sqrt{q}h_c(0)<1/\sqrt{K}$. Because
\begin{equation}\label{equ:69}
\begin{split}
  &2\sqrt{q}h_c(0)-\frac{1}{\sqrt{K}}\\
=&\frac{1}{\sqrt{2q}}\int_{-1}^1(2s_\omega-1)[\frac{\omega^2}{q\sqrt{q^2-\omega^2}}\frac{1}{\sqrt{1+\sqrt{1-\omega^2/q^2}}}\\
&-\sqrt{1+\sqrt{1-\omega^2/q^2}}]g(\omega)d\omega.
\end{split}
\end{equation}
Straightforward calculation shows that a negative value of Eq.~(\ref{equ:69}) is equivalent to $F'(q)<0$ for the self-consistent equation (\ref{equ:20}). Otherwise, a positive root implies $2\sqrt{q}h_c(0)-1/\sqrt{K}>0$, which is equivalent to $F'(q)>0$.

To obtain analytical insights, we take frequency distribution to be a bimodal Dirac, i.e., $g(\omega)=\frac{1}{2}[\delta(\omega-1)+\delta(\omega+1)]$. Now $q_l=1$ and the parametrical function
\begin{equation}\label{equ:70}
  F(q)=(2s-1)\sqrt{\frac{1+\sqrt{1-q^{-2}}}{2q}}.
\end{equation}
Clearly, $q_c=2/\sqrt{3}$ such that $F'(q_c)=0$. $F'(q)>0$ for $q\in [q_l, q_c)$ and $F'(q)<0$ for $q\in(q_c, \infty)$. The nonzero eigenvalue $\lambda$ satisfying Eq.~(\ref{equ:64}) is obtained as
\begin{equation}\label{equ:71}
  \lambda=\frac{2}{q(1+\sqrt{1-q^{-2}})}-2\sqrt{q^2-1}.
\end{equation}
It is easily verified that $\lambda$ is positive for $q\in[q_l, q_c)$ and is negative for $q\in(q_c, \infty)$, which coincides with the argument above.

Turning briefly to the incoherent state, where no oscillators are phase locked and $R_m=q=0$, then the operator $\mathcal{B}$ is $\mathbf{0}$, $\mathcal{L}=\mathcal{M}$, and $\sigma_p(\mathcal{L})$ is empty. The spectrum of $\mathcal{L}$ contains the continuous spectrum which is the entire imaginary axis.

\begin{figure}
  \centering
  \includegraphics[width=\linewidth]{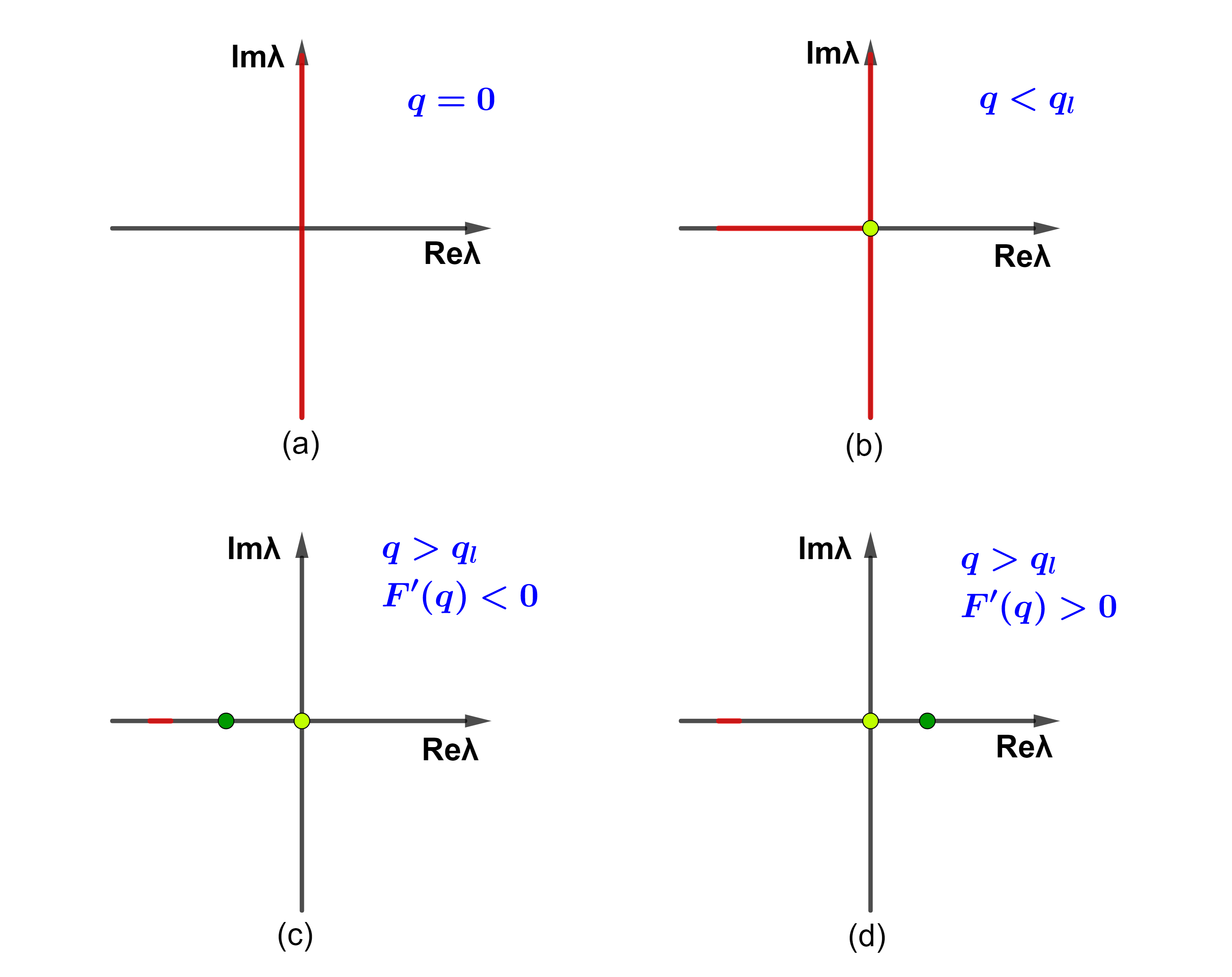}\\
  \caption{Diagrammatic sketch of the spectrum of linear operator. (a) the incoherent state (neutral stable), (b) the partially locked state (linearly neutral stable), (c) the fully locked state (linearly stable), (d) the fully locked state (linearly unstable). The red lines on the imaginary axis and real axis represent the continuous spectrum formed by the drifting and the locked oscillators respectively. The light green point at the origin is the trivial eigenvalue of the linear operator. The dark green point on the real axis represents the nontrivial eigenvalue of the linear operator.}\label{fig:01}
\end{figure}

We now present a brief summary of the spectrum of $\mathcal{L}$ for all steady states in the infinite $N$ Kuramoto model with higher-order interactions. For the incoherent state, there are no frequency locked oscillators, and the operator $\mathcal{L}=\mathcal{M}$, which has purely continuous spectrum $i\mathbb{R}$ on the imaginary axis [Fig.~\ref{fig:01}(a)]. This spectrum structure rules out any kind of linearly stable or unstable mode in the tangent space, and thus the incoherent state is neutrally stable to perturbation for any value of $K$. One can never expect it to lose stability at the critical point, at which the partially synchronized states are born. For the partially locked state with $q<q_l$, where the drifting oscillators and the locked ones coexist. The continuous spectrum exhibits an inverted $T$-shape that includes the entire imaginary axis along with the segment $[-2q, 0]$ [Fig.~\ref{fig:01}(b)]. Furthermore, there are no other roots to the characteristic equations except for a trivial eigenvalue $\lambda=0$ originating from the rotational symmetry of the system. In this sense, we call the partially locked case linearly neutral stable, due to the presence of negative value of $\sigma_c(\mathcal{L})$. For the fully locked state with $q>q_l$, there are no drifting oscillators. The continuous spectrum consists of the closed interval $[-2q, -2\sqrt{q^2-1}]$. In addition to a trivial eigenvalue $\lambda=0$, a negative eigenvalue exists in $(-2\sqrt{q^2-1}, 0)$ corresponding to $F'(q)<0$ [Fig.~\ref{fig:01}(c)], and the positive eigenvalue appears corresponding to $F'(q)>0$ [Fig.~\ref{fig:01}(d)]. This suggests that the fully locked state is linearly stable [$F'(q)<0$] or unstable [$F'(q)>0$] signaling a saddle node bifurcation for $q$ passing through $q_c$.

\section{Conclusion}\label{sec:05}

In this paper, we have systematically studied the dynamical properties of steady states in the Kuramoto model with higher-order interactions in the thermodynamic limit. The phase model serves as a representative nonlinear higher-order coupling that has been shown to exhibit novel collective dynamics for phase transitions to synchronization. Based on the self-consistent analysis in continuous limit $N\rightarrow\infty$, all steady states are well characterized and the underlying bifurcation can be detected from the parametrical function with a single variable. Importantly, we provide a complete understanding of the spectrum of fixed states by linearizing the continuity equation. We demonstrate that the unperturbed linear operator contains only the continuous spectrum, which is the entire imaginary axis or a closed interval on the negative real axis corresponding to the drifting and locked oscillators respectively. Furthermore, we prove that the drifting oscillators do not contribute to neither the order parameters nor the characteristic functions, but they have significant influence on the structure of continuous spectrum on the real axis. According to whether the nontrivial eigenvalue of the linear operator exists or not, we argue that the incoherent state, the partially locked state and the fully locked state are respectively neutral stable, linearly neutral stable, and linearly stable (unstable) to perturbation in tangent space.

\section*{ACKNOWLEDGMENTS}
This work is supported by the National Natural Science Foundation of China (Grants No. 11905068) and the Scientific Research Funds of Huaqiao University (Grant No. ZQN-810).


\begin{thebibliography}{99}
%introduction
\bibitem{strogatz2003sync} S.H. Strogatz, \emph{Sync: The Emerging Science of Spontaneous Order} (Hypernion, New York, 2003).
\bibitem{Pikovsky2003} A. Pikovsky, M. Rosenblum, and J. Kurths, {\it Synchronization: a universal concept in nonlinear sciences} (Cambridge University Press, 2003).
\bibitem{Glass1988} L. Glass and M. C. Mackey, {\it From Clocks to Chaos: The Rhythms of Life} (Princeton University Press, Princeton, 1988).
\bibitem{benz1991coh}S. Benz and C. Burroughs, Coherent emission from two-dimensional Josephson junction arrays, Appl. Phys. Lett. {\bf 58}, 2162 (1991).
\bibitem{Rohden2012PRL} M. Rohen, A. Sorge, M. Timme, and D. Witthaut, Self-organized synchronization in decentralized power grids, Phys. Rev. Lett. {\bf 109}, 064101 (2012).
\bibitem{prindle2012sen}A. Prindle, P. Samayoa, I. Razinkov, T. Danino, L.S. Tsimring, and J. Hasty, A sensing array of radically coupled genetic 'biopixels', Nature (London) {\bf 481}, 39 (2012).\bibitem{buzski2004sci}G. Buzs{\'a}ki and A. Draguhn, Neuronal Oscillations in Cortical Networks, Science {\bf 304}, 1926 (2004).
\bibitem{Arenas2008PhysRep} A. Arenas, A.  D\'{i}az-Guilera,
  J. Kurths, Y. Moreno, C. Zhou, Synchronization in complex
  networks. {\it Phys. Rep.} {\bf 469}, 93--153 (2008).
\bibitem{kuramoto1975}Y. Kuramoto, in \emph{International Symposium on Mathematical Problems in Theoretical Physics}, edited by H. Araki, Lecture Notes in Physics No. 30 (Springer, New York, 1975), p. 420.
\bibitem{strogatz2000from}S. H. Strogatz, From Kuramoto to Crawford: exploring the onset of synchronization in populations of coupled oscillators. Phys. D {\bf 143}, 1 (2000).
\bibitem{acebron2005the} J.A. Acebr{\'o}n, L.L. Bonilla, C.J. P{\'e}rez Vicente, F. Ritort, and R. Spigler, The Kuramoto model: A simple paradigm for synchronization phenomena. Rev. Mod. Phys. \textbf{77}, 137 (2005).
\bibitem{pikovsky2015dyn}A. Pikovsky, and  M. Rosenblum, Dynamics of globally coupled oscillators: Progress and perspectives. Chaos: An Interdisciplinary Journal of Nonlinear Science, {\bf 25}(9), 097616 (2015).
\bibitem{ott2008low}E. Ott and T. M. Antonsen, Low dimensional behavior of large systems of globally coupled oscillators, Chaos {\bf 18}, 037113 (2008).
\bibitem{ott2009long}E. Ott and T. M. Antonsen, Long time evolution of phase oscillator systems, Chaos {\bf 19}, 023117 (2009).
\bibitem{Ashwin2016PhysD} P. Ashwin and Ana Rodrigues, Hopf normal form with $S_N$ symmetry and reduction to systems of nonlinearly coupled phase oscillators, Physica D {\bf 325}, 14 (2016).
\bibitem{Leon2019PRE} I. Le\'{o}n and D. Paz\'{o}, Phase reduction beyond the first order: The case of the mean-field complex Ginzburg-Landau equation, Phys. Rev. E {\bf 100}, 012211 (2019).
\bibitem{Petri2014Interface} G. Petri, P. Expert, F. Turkheimer, R. Carhart-Harris, D. Nutt, P.J. Hellyer, and F. Vaccarino, Homological scaffolds of brain functional networks, J. R. Soc. Interface {\bf 11}, 20140873 (2014).
\bibitem{Giusti2016JCN} C, Giusti, R. Ghrist, and D. S. Bassett, Two's company, three (or more) is a simplex, J. Comput. Neurosci. {\bf 41}, 1 (2016).
\bibitem{Sizemore2018JCN} A. E. Sizemore, C. Giusti, A. Kahn, J. M. Vettel, R. Betzel, and D. S. Bassett, Cliques and cavities in the human connectome, J. Comput. Neurosci. {\bf 44}, 115 (2018).
\bibitem{reimann2017cliques}M.W. Reimann, M. Nolte, M. Scolamiero, K. Turner, R. Perin, G. Chindemi, P. Dlotko, R. Levi, K. Hess, and H.Markram, Cliques of neurons bound in to cavities provide a missing link between structure and function, Front. Comput. Neurosci. \textbf{11}, 48 (2017).
\bibitem{Battiston2020PhysRep} F. Battiston, G. Cencetti, I. Iacopini, V. Latora, M. Lucas, A. Patania, J.-G. Young, and G. Petri, Networks beyond pairwise interactions: Structure and dynamics, Phys. Rep. {\bf 874}, 1 (2020).
\bibitem{komarov2015finite}M. Komarov, and A. Pikovsky, Finite-size-induced transitions to synchrony in oscillator ensembles with nonlinear global coupling, Phys. Rev. E \textbf{92}, 020901(R) (2015)
\bibitem{bibk2016chaos}C. Bick, P. Ashwin, and A. Rodrigues, Chaos in generically coupled phase oscillator networks with nonpairwise interactions, Chaos \textbf{26}, 094814 (2016).
\bibitem{gong2019low}C.C. Gong, and A. Pikovsky, Low-dimensional dynamics for higher order harmonic globally coupled phase oscillator ensemble, Phys. Rev. E {\bf 100}, 062210 (2019).
\bibitem{salnikov2019simplicial} V. Salnikov, D. Cassese, and R. Lambiotte, Simplicial complexes and complex systems, Eur. J. Phys. \textbf{40}, 014001 (2019).
\bibitem{skardal2011cluster}P.S. Skardal, E. Ott, and J.G. Restrepo, Cluster synchrony in systems of coupled phase oscillators with higher-order coupling, Phys. Rev. E {\bf 84}, 036208 (2011).
\bibitem{komarov2013multi} M. Komarov and A. Pikovsky, Multiplicity of Singular Synchronous States in the Kuramoto Model of Coupled Oscillators, Phys. Rev. Lett. {\bf 111}, 204101 (2013).
\bibitem{xu2016col}C. Xu, H. Xiang, J. Gao, and Z. Zheng. Collective dynamics of identical phase oscillators with high-order coupling. Scientific Reports, {\bf 6}, 31133 (2016).
\bibitem{wang2017syn}H. Wang, W. Han, and J. Yang. Synchronous dynamics in the Kuramoto model with biharmonic interaction and
bimodal frequency distribution. Phys. Rev. E {\bf 96}, 022202 (2017).
\bibitem{filatrella2007gen}G. Filatrella, N. F. Pedersen, and K. Wiesenfeld, Generalized coupling in the Kuramoto model, Phys. Rev. E {\bf 75}, 017201 (2007).
\bibitem{zou2020dyn}W. Zou, and  J. Wang, Dynamics of the generalized Kuramoto model with nonlinear coupling: Bifurcation and stability. Phys. Rev. E {\bf 102}, 012219 (2020).
\bibitem{Millan2020PRL} A. P. Mill\'{a}n, J. J. Torres, and G. Bianconi, Explosive higher-order Kuramoto dynamics on simplicial complexes, Phys. Rev. Lett. {\bf 124}, 218301 (2020).
\bibitem{Skardal2020} P.S. Skardal and A. Arenas, Higher-order interactions in complex networks of phase oscillators promote abrupt synchronization switching, Commun. Phys. {\bf 3}, 218 (2020).
\bibitem{Skardal2020b}P.S. Skardal and A. Arenas, Memory selection and information switching in oscillator networks with higher-order interactions, J. Phys. Complex. {\bf 2}, 015003 (2020).
\bibitem{Lucas2020PRR} M. Lucas, G. Cencetti, and F. Battiston, Multiorder Laplacian for synchronization in higher-order networks, Phys. Rev. Res. {\bf 2}, 033410 (2020).
\bibitem{skardal2019abrupt}P.S. Skardal, and A. Arenas, Abrupt Desynchronization and Extensive Multistability in Globally Coupled Oscillator Simplexes, Phys. Rev. Lett. {\bf 122}, 248301 (2019).
\bibitem{xu2020bif}C. Xu, X. Wang, and P. S. Skardal, Bifurcation analysis and structural stability of simplicial oscillator populations, Phy. Rev. Research {\bf 2}, 023281 (2020).

%context

\bibitem{strogatz1991stability} S.H. Strogatz, and R.E. Mirollo, Stability of incoherence in a population of coupled oscillators, J. Stat. Phys. {\bf 63}, 613 (1991).
\bibitem{omel2013bifurcation}O.E. Omel'chenko, and M. Wolfrum, Bifurcations in the SakaguchišCKuramoto model, Physica D {\bf 263}, 74-85 (2013).
\bibitem{mirollo2007spe}R. Mirollo, and S. H. Strogatz, The Spectrum of the Partially Locked State for the Kuramoto Model, J. Nonlinear Sci. {\bf 17}, 309-347 (2007).
\bibitem{mirollo2005spe}R.E. Mirollo, and S.H. Strogatz, The spectrum of the locked state for the Kuramoto model of coupled oscillators, Physica D \textbf{205}, 249-266 (2005).

\end{thebibliography}
\end{document}